\documentclass[preprint,pra,superscriptaddress,aps,showpacs,showkeys,amsmath,floatfix]{revtex4-1}

\usepackage{graphics}
\usepackage{graphicx}
\usepackage{epstopdf}
\usepackage{dcolumn}
\usepackage{bm}
\usepackage{longtable}
\usepackage{epsfig}
\usepackage{times}
\usepackage{url}
\usepackage{color}

\begin{document}
%%%%%%%%%%%%%%%%%%%%%%%%%%%%%%%%%%%%%%%%%%%%%%%%%%%%%%%%%%%%%%%%%%%%%%%%%%%%%%%%
%%%%%%%%%%%%%%%%%%%%%%%%%%%%%%%%%%%%%%%%%%%%%%%%%%%%%%%%%%%%%%%%%%%%%%%%%%%%%%%%
\title{Transient nuclear inversion by X-Ray Free Electron Laser in a tapered x-ray waveguide}

\author{Yu-Hsueh   \surname{Chen}}
\affiliation{Department of Physics, National Central University, Taoyuan City 32001, Taiwan}
\affiliation{Max-Planck-Institut f\"ur Kernphysik, Saupfercheckweg 1, D-69117 Heidelberg, Germany}

\author{Po-Han   \surname{Lin}}
\affiliation{Department of Physics, National Central University, Taoyuan City 32001, Taiwan}

\author{Guan-Ying   \surname{Wang}}
\affiliation{Department of Physics, National Central University, Taoyuan City 32001, Taiwan}

\author{Adriana \surname{P\'alffy}}
\email{adriana.palffy-buss.fau.de}
\affiliation{Department of Physics, Friedrich-Alexander-Universit\"at Erlangen-N\"urnberg, D-91058, Erlangen, Germany } 
\affiliation{Max-Planck-Institut f\"ur Kernphysik, Saupfercheckweg 1, D-69117 Heidelberg, Germany}

\author{Wen-Te \surname{Liao}}
\email{wente.liao@g.ncu.edu.tw}
\affiliation{Department of Physics, National Central University, Taoyuan City 32001, Taiwan}
\affiliation{Max-Planck-Institut f\"ur Kernphysik, Saupfercheckweg 1, D-69117 Heidelberg, Germany}
\affiliation{Physics Division, National Center for Theoretical Sciences, Hsinchu 30013, Taiwan}
\date{\today}
%%%%%%%%%%%%%%%%%%%%%%%%%%%%%%%%%%%%%%%%%%%%%%%%%%%%%%%%%%%%%%%%%%%%%%%%%%%%%%%%
%%%%%%%%%%%%%%%%%%%%%%%%%%%%%%%%%%%%%%%%%%%%%%%%%%%%%%%%%%%%%%%%%%%%%%%%%%%%%%%%
\begin{abstract}
By restricting the spatial energy transmission of an electromagnetic wave, dielectric waveguides transmit light over long distances at sustained intensity. Waveguides have been used in the microwave and optical range to maintain strong signal intensities in connection with lasers, but guiding of intense short-wavelength radiation such as x-rays has proven more cumbersome. 
Here we investigate theoretically how tapered  x-ray waveguides can focus and guide radiation from  x-ray free electron lasers.  Elliptical waveguides using a cladding material with high atomic number such as platinum can maintain an x-ray intensity up to three orders of magnitude larger than in free space. This feature can be used to significantly  enhance resonant interactions of x-rays, for instance driving nuclear transitions up to transient nuclear population inversion. This could be the first breakthrough in nuclear state population control. Our results anticipate the important role of tapered x-ray waveguides in the emerging field of x-ray quantum optics with nuclear transitions. 
\end{abstract}
%%%%%%%%%%%%%%%%%%%%%%%%%%%%%%%%%%%%%%%%%%%%%%%%%%%%%%%%%%%%%%%%%%%%%%%%%%%%%%%%

\keywords{x-ray quantum optics, x-ray free electron laser, waveguides, nuclear quantum optics}
%%%%%%%%%%%%%%%%%%%%%%%%%%%%%%%%%%%%%%%%%%%%%%%%%%%%%%%%%%%%%%%%%%%%%%%%%%%%%%%%
\maketitle
%%%%%%%%%%%%%%%%%%%%%%%%%%%%%%%%%%%%%%%%%%%%%%%%%%%%%%%%%%%%%%%%%%%%%%%%%%%%%%%%
%-----------Text body-----------------------

%%%%%%%%%%%%%%%%%%%%%%%%%%%%%%%%%%%%%%%%%%%%%%%%%%%%%%%%%%%%%%%%%%%%%%%%%
More than sixty years ago, it was the invention of the laser that revolutionized atomic physics, leading to a better understanding of atomic and molecular dynamics and to a vast number of applications in both fundamental and applied sciences. With only optical frequencies available, the interaction of coherent light with matter was for a long time mainly restricted to driving and control of atomic  transitions. Fortunately, since 2009 we have witnessed the commissioning and operation of the first x-ray free electron laser (XFEL) facilities \cite{Emma2010.NP,Pellegrini2016}, opening for study the interaction of high-frequency lasers with matter.  X-ray photons carry large momentum, are easy to detect, can be focused to much smaller spot sizes, can penetrate more deeply into materials and, in principle, can support faster information processing because of their higher frequencies \cite{OlgaNews2017}. As such, they would be very desirable as information carriers for quantum technologies, for exploring the boundary between the classical and quantum world \cite{Haroche2001}, or for matter-probing techniques based on quantum effects.  

Quantum optics with visible and infrared photons relies on resonant interactions of photons with atoms \cite{Haroche2001}. In contrast, with their high frequencies, x-rays are no longer resonant to valence electron transitions in atoms, but rather with inner-shell transitions in highly charged ions, or even with keV electromagnetic transitions in the atomic nucleus \cite{Adams2013}. An advantage of interactions based on nuclear over electronic transitions is that nuclear transitions have narrow spectral linewidths even in bulk solids at room temperature. This renders them very clean quantum systems, with long coherence times. These desirable properties are due to the large mass and small size of nuclei, as well as the possibility for recoilless absorption of gamma rays and x-rays (the M\"ossbauer effect) \cite{OlgaNews2017,Olga2014}. The downside is that due to the tiny nuclear size, the light-nuclear coupling is typically five orders of magnitude weaker  than the electromagnetic coupling of any atomic system.
Thus, x-ray quantum optics experiments  have been successfully performed at the single-photon level with a very narrow  M\"ossbauer nuclear transition at 14.4 keV in the stable  $^{57}$Fe nucleus \cite{Roehlsberger2010,Roehlsberger2012,Heeg2013,Adams2013,Olga2014,Heeg2015a,Heeg2015b,Heeg2017,Haber2017,Chumakov2018,Heeg2021}.  
However, even intense XFEL radiation is still far from achieving efficient nuclear pumping or nuclear population inversion \cite{Chumakov2018}. Apart from x-ray quantum optics incentives, efficient driving of nuclear transitions and nuclear population control would be desirable for a safe and clean energy storage solution based on nuclear isomers, i.e., long-lived nuclear excited states \cite{Walker1999}.

An intuitive solution to this probem is given by x-ray nano-focusing, either in free space on in specially designed structures. At present, tight x-ray nano-focusing optics is yet to be developed at XFEL facilities \cite{LCLSBLs, SACLABLs, EuropeanXFELBLs, SwissFELBLs, PalFELBLs}. Thus, 
x-ray and nuclear quantum optics applications would very much benefit from the development of structures that enhance the light-matter interactions such as waveguides or cavities. So far,  grazing incidence x-ray cavities with embedded M\"ossbauer nuclei \cite{Roehlsberger2010,Roehlsberger2012,Heeg2013,Heeg2015a,Heeg2015b,Haber2016, Haber2017} or normal-incidence cavities based on diamond mirrors \cite{Shvydko2010,Shvydko2011} have been shown to be very lossy. Theoretical prospects of strong nuclear excitation in x-ray thin-film cavities were discussed in Ref.~\cite{Heeg-arXiv2016}. 
 X-ray waveguides have been designed and fabricated to work with x-ray radiation from synchrotron sources \cite{Pfeiffer2002, Jarre2005,Kruger2012,Chen2015,Hoffmann2016}. Such waveguides have been successfully used as point-like hard x-ray sources for imaging and in particular for holographical imaging \cite{Salditt2015}.

\begin{figure}[b]
\includegraphics[width=0.7\textwidth]{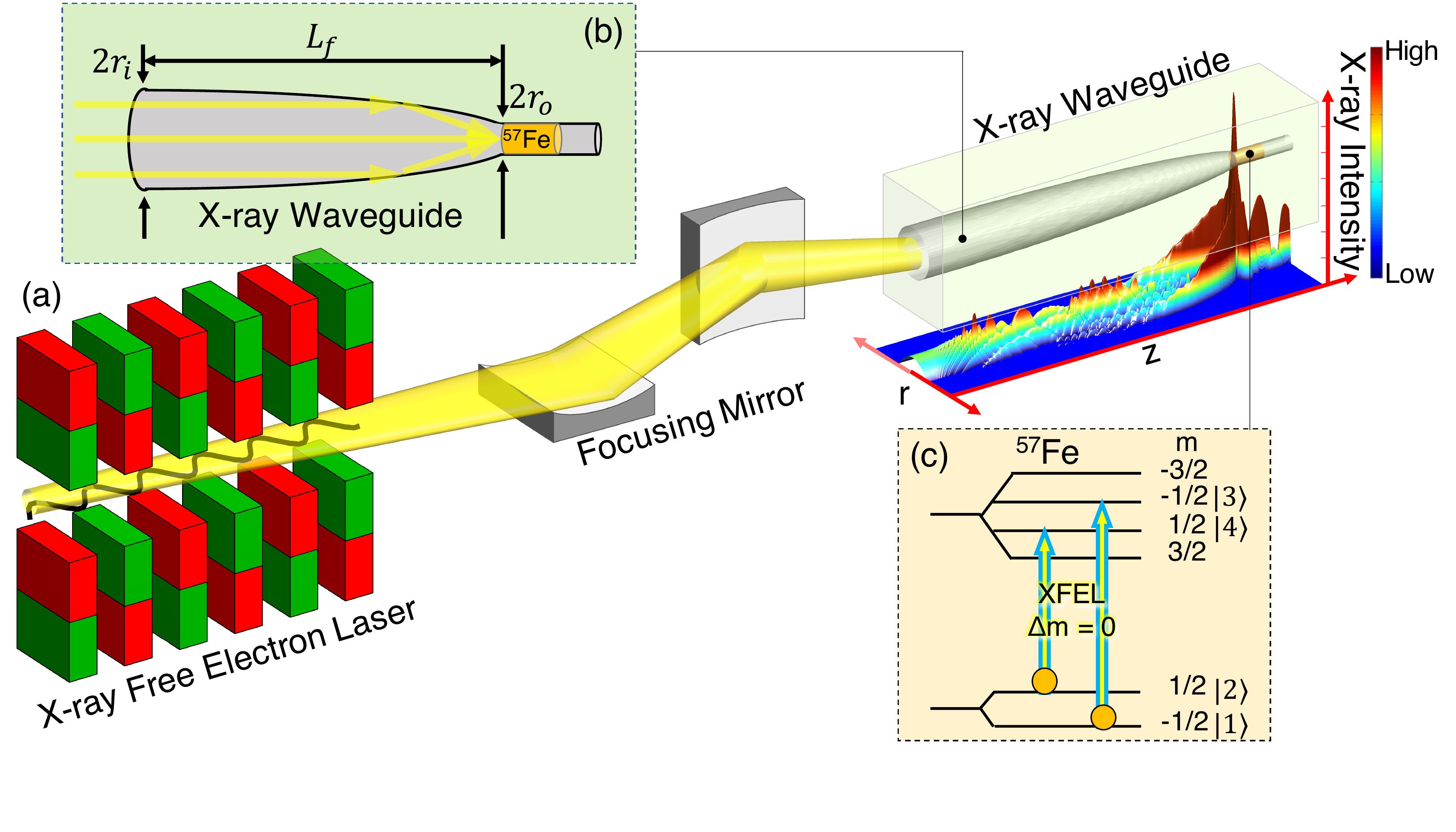}
\caption{\label{fig1} 
{\bf Nuclear sample embedded in an X-ray waveguide.} 
(a) XFEL radiation is focused on the tapered WG. A nuclear sample (yellow cylinder) is placed at the focal point of the WG (sketch not to scale). The x-ray intensity in the waveguide is illustrated below. 
(b) A tapered WG has a bottle-like shape, with input radius $r_i$, output radius $r_o$ and focusing length $L_f$. 
(c)  $^{57}$Fe nuclear level scheme. A linearly polarized XFEL pulse (blue arrows) drives the two $\Delta m=0$ transitions. We label the two ground states with $|1\rangle$, $|2\rangle$ and the relevant excited states with $|3\rangle$ and $|4\rangle$, respectively.
}
\end{figure}

Here, we investigate for the first time the theoretical prospects of tapered x-ray waveguides (WG) for efficient nuclear pumping of a nuclear transition by an XFEL. 
A tapered WG presents a gradual narrowing structure design of the cladding material.  Such a WG not only confines but also focuses  x-rays \cite{Stern1988, Denisov1991, Bergemann2003, Kruger2012, Chen2015}. This boosts the x-ray capability of pumping nuclear excitations. An additional important advantage of the WG is that nuclear resonance fluorescence therein will proceed by photons emitted in designed  modes, leading to  directional emission even without the generation of a superradiant state \cite{Roehlsberger2010}.   This is relevant for downstream detection and applications. The waveguide setup is illustrated in Fig.~\ref{fig1}(a). A nuclear sample  is embedded around the focal point of an X-ray cladding waveguide (WG) depicted in  Fig.~\ref{fig1}(a) by the gray bottle-shaped volume.  
The typical parameters describing a tapered WG are the  focusing length $L_f$, the input (output) port radius $r_i$ ($r_o$), the cladding material, and the cladding-vacuum interface geometry illustrated in Fig.~\ref{fig1}(b). 
As a main example we consider the M\"ossbauer nuclear transition from the stable $^{57}$Fe ground state to the first excited nuclear state at 14.4 keV. This corresponds to an x-ray wavelength of approx. 1 {\AA} and has magnetic dipole multipolarity. Similar transitions in  $^{133}$Ba,  $^{187}$Os, and $^{169}$Tm that will be also addressed are presented in Table \ref{table1}.  

%

%%%%%%%%%%%%%%%%%%%%%%%%%%
%%%%%%%%%%%%%%%%%%%%%%%%%

In the theoretical modelling we consider in the following the XFEL pulse to be linearly polarized and fully coherent. The effects of limited coherence will be modeled and discussed separately in the Results. Fully coherent x-ray pulses can be delivered by seeded XFELs \cite{Feldhaus1997, Amann2012, Serkez2016, Emma2017,Inoue2019,Nam2021}, 
an XFEL oscillator (XFELO) \cite{Kim2008, Zemella2012, Aadams2019, Shvydko2019, Rauer2019,Marcus2020}, or their combination \cite{Li2018, Paraskaki2019}. We would like to highlight here the very recent report of successful seeding at the Pohang Accelerator Laboratory XFEL in South Korea, which demonstrated 40 times higher peak brightness than in self-amplified spontaneous emission, with pulses close to the Fourier limit and energies up to 14.6 keV \cite{Nam2021}. This energy range would cover all examples of nuclear transitions discussed here. 
 The nuclear transitions usually experience hyperfine splitting according to their respective spins. For all four considered transitions, $I_u=3/2$ and $I_d=1/2$  are the nuclear spins of the ground (down, $d$) and excited (up, $u$) states, respectively.  We consider a setup geometry which allows the XFEL pulse to drive the two transitions $\Delta m=m_u-m_d=0$, where $m_u$ ($m_d$) are the spin projections of the excited (ground) nuclear state on the quantization axis.  The dynamics of the system can be modelled theoretically  by the so-called Maxwell-Bloch equations \cite{Scully2006, Agarwal2012} which combine the dynamics of the density matrix of the nuclear system with the Maxwell equation describing the coherent field propagation. The nuclear dynamics can be expressed in the Liouville equation \cite{Buervenich2006, Liao2011, Liao2012a, Liao2012b, Liao2013, Kong2014, Kong2016, Wang2018}
\begin{equation}
	\dot{\hat{\rho}}=-\frac{i}{\hbar}\left[\hat{H},\hat{\rho}\right]+\mathcal{L}[\hat{\rho}]\, ,
	\label{Liouv}
\end{equation}
where $\hat{\rho}$ is the density matrix describing the nuclear states, $\hbar$ stands for the reduced Planck constant, and $\hat{H}$ is the Hamiltionian of the system. Furthermore, the losses in the system are described by the Lindblad superoperator $\mathcal{L}[\hat{\rho}]$. The Maxwell wave equation describing the coherent propagation of the x-ray electric field modulus $E=|\vec{E}|$ through the medium is given by 
\cite{Shvydko1998, Fuhse2006, Liao2012a, Kong2014}
\begin{eqnarray}\label{eq4}
 &  & \left( \partial_x^2 +\partial_y^2 \right)E +2ik \left( \partial_{z} + \frac{1}{c}\partial_{t} \right)  E + k^2 \left[ n\left( \vec{r} \right)^2-1\right] E \nonumber\\
 &=&  - \left[\frac{8\pi\hbar\Gamma_s N\left( \vec{r} \right) }{\Pi k}\frac{2I_u+1}{2I_d+1}\frac{f_{lm}}{\alpha+1} \right]\left(  \rho_{31}+\rho_{42} \right) .
\end{eqnarray}
Here, $k$ is the x-ray wave vector and $c$ denotes the speed of light in vacuum. The complex x-ray index of refraction determined by the electrons in the sample is denoted by  $n\left( \vec{r} \right)=1-\delta\left( \vec{r} \right)+i \beta\left( \vec{r} \right)$ \cite{Fuhse2006, Bukreeva2006, Chen2015}, with the values 
for different materials presented in Table \ref{table1}. Furthermore, $\Gamma_s$ is the spontaneous decay rate  of the nuclear excited state, $N\left( \vec{r}\right)$ the nuclear particle number density distribution, $f_{lm}$ is the Lamb-M\"ossbauer factor, and  $\alpha$ is the internal conversion coefficient, respectively. 
The notation $\Pi$ defined in the Methods stands for the nuclear matrix element and  $\rho_{31}$ and $\rho_{42}$ are the off-diagonal density matrix elements for the nuclear-four-level  scheme depicted in Fig.~\ref{fig1}(c).
The two coupled equations \eqref{Liouv} and \eqref{eq4} are solved numerically in cylindrical coordinates $(r,z)$ (see Methods). 

\section{Results}

%%%%%%%%%%%%%%%%%%%%%%%%%%%%%%%%%%%%%%%%
\begin{figure}[b]
\includegraphics[width=0.6\textwidth]{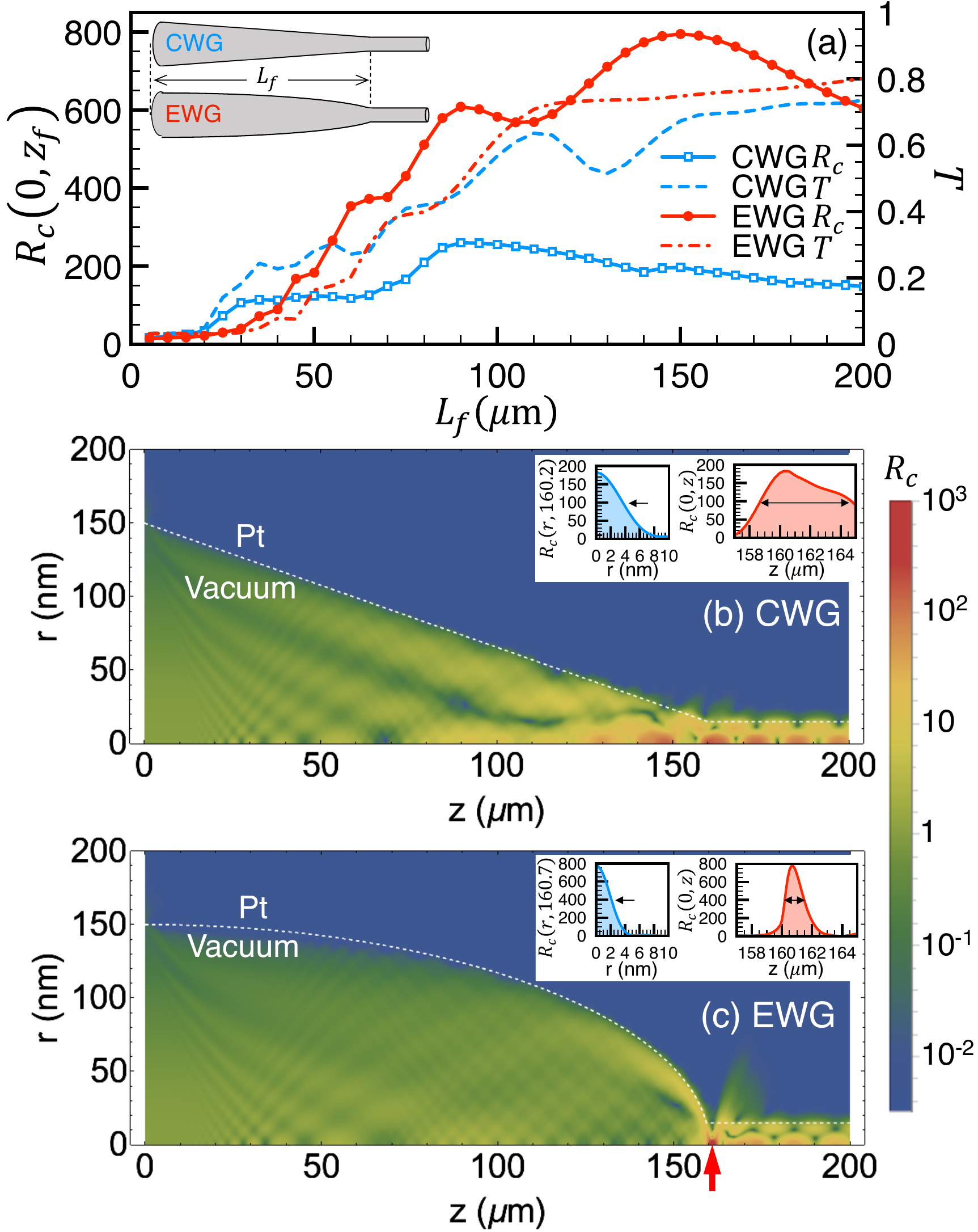}
\caption{\label{fig2} 
{\bf Focusing X-rays in a tapered waveguide.} 
(a) Intrawaveguide x-ray concentration ratio $R_c(r=0,z)$ (left vertical axis) and  transmission $T$ (right vertical axis) of the CWG (blue empty squares and blue-dashed line) and EWG (red  dots and red-dashed-dotted line). For comparison we consider both WG with ($r_i$,  $r_o$) = (150~nm, 15~nm).
(b, c) Contour plots of x-ray concentration distribution $R_c\left( r, z\right) $  for a Pt-cladding (b) CWG and (c) EWG. The common WG parameters are  ($r_i$, $L_f$, $r_o$) = (150~nm, 160~$\mu$m, 15~nm). The input x-ray pulse has a  transverse  Gaussian form with  $\sigma =$120~nm. 
The red-upward arrow pinpoints the EWG focal point.
White-dashed lines indicate the platinum cladding and vacuum interface. 
Insets in \mbox{(b, c)} illustrate radial  and axial $R_c$ values about the WG focal point.  The  black horizontal arrows indicate the focal spot size (left) and intrawaveguide Rayleigh length (right).
}
\end{figure}
%%%%%%%%%%%%%%%%%%%%%%%%%%%%%%%%%%%%%%%%%

We  first  characterize the quality of the x-ray waveguide. To this end we define and calculate the 
 intrawaveguide x-ray concentration ratio $R_c$ and the transmission $T$
\begin{eqnarray}
R_c \left( r, z\right)& = & \frac{\vert E\left( r, z\right)  \vert^2}{\vert E\left(  0, 0 \right)  \vert^2}, \label{eq7}\\
T & = & \frac{\int_0^{r_o} \vert E\left(  r, z_f \right)  \vert^2 r dr}{\int_0^{r_i} \vert E\left( r, 0 \right)  \vert^2 r dr}, \label{eq8}
\end{eqnarray}
for different waveguide geometries.  Here $z = z_f$ is the axial position where the maximum intrawaveguide x-ray intensity occurs.
From their definitions it is clear that  $R_c$ depicts the degree of x-ray focusing and the spatial intensity distribution, while $T$ reveals the percentage of transmitted (and therefore not lost) x rays in a WG. In Fig.~\ref{fig2} we present our steady-state numerical solutions of equation \eqref{eq4} with only the electronic contribution of the cladding index of refraction (see Table~\ref{table1}).   
The photon loss usually accompanies the reflection on the cladding-vacuum interface. 
Intuitively, an elliptical waveguide (EWG), whose interface is elliptical, will guide x rays from one focal point to another via only a single reflection on the interface. One can therefore take this advantage to reduce photon loss when channelling XFEL through a EWG. In order to show this advantage, we compare the EWG with a conical waveguide (CWG), whose cladding-vacuum interface is a cone.

We consider an EWG and a CWG, both sharing the geometrical parameters ($ r_i,  r_o$) = (150~nm,  15~nm),  illuminated by the same input XFEL of beam waist $\sigma =$120~nm at WG input port.  X-rays experience focusing for $z \leq 160\ \mu$m in both WGs, but behave very differently for $z > 160\ \mu$m. 
By varying the $L_f$, Fig.~\ref{fig2}(a) demonstrates the performance of the two types of tapered WG and shows the $L_f$-dependent axial x-ray concentration $R_c \left( 0, z_f\right)$  and  the transmission $T$.
We notice that the performance of EWG is generally better than that of CWG.
The  EWG transmission becomes higher than that of CWG when $L_f \geq 80\ \mu$m, and the gap between the two $R_c$ curves is increasing for $L_f > 40\ \mu$m. 
In contrast to the maximum CWG $R_c \approx 250$ at $L_f = 90\ \mu$m, EWG $R_C$ remarkably reaches the greatest concentration ratio of  approx.~800 at $L_f = 150\ \mu$m. 
The product $T R_c \left( 0, z_f\right)$ can be used as the WG figure of merit. The EWG's  value reaches the optimized case around $L_f = 160\ \mu$m,  where it offers both tight X-ray focusing and high transmission.

In order to demonstrate a detailed comparison between the two geometries, Figs.~\ref{fig2}(b-c) illustrate the spatial distribution of $R_c \left( r, z\right)$ for Pt-cladding CWG and EWG at $ L_f$ = 160~$\mu$m where the  EWG is optimized.  The axial dependence  defines the intrawaveguide Rayleigh length as the  full width half maximum (FWHM) of $R(0,z)$.  
The EWG $R_c$ value is four times that of CWG at the maximal point $z_{\rm max} \approx 161~\mu$m where $R_c \left( 0, z_\mathrm{max}\right)\approx 800$ (see red arrow in Fig.~\ref{fig2}c).
The EWG is capable of better focusing and maintaining higher transmission than the CWG. 
Although the CWG provides looser focusing compared to the EWG, its more uniform field distribution  may render it a good platform for observing collective emission \cite{Roehlsberger2010}. For the present purpose of producing as strong a nuclear excitation as possible, we will focus on EWG in the following.

\noindent
{\bf XFEL-pumped transient nuclear inversion in an elliptical waveguide.}
%%%%%%%%%%%%%%%%%%%%%%%%%%%%%%%%%%%%%%%%%%%%%%%%%%%%%%%%%%%%%%%%%%%%%%%%%%%%%%%
Let us now investigate the efficiency of nuclear excitation in an EWG. To this end we numerically solve Eqs.~(\ref{Liouv}) and (\ref{eq4}) (see Methods). The nuclear inversion is defined as   $I_v = \rho_{33}+\rho_{44} -\rho_{11}-\rho_{22}$ \cite{Buervenich2006} (see Methods). We can  express it  with the help of the x-ray concentration $R_c(r,z)$  and further pulse parameters such as the beam waist $\sigma$ and the number of photons in the pulse $n_p$ (see Methods). The required maximum XFEL spot radius $\sigma_{\mathrm{max}}$ for achieving full nuclear inversion $I_v=1$ can be estimated  via Eq.~(\ref{eq9}). Considering $R_c=1$, i.e., no waveguide focusing effects, 
pulse duration   $\tau = 42.5$~fs corresponding to the FWHM  $2 \sqrt{2\ln 2} \tau =100$~fs  of the European XFEL \cite{Geloni2010},  
and the values  of isotopes listed in Table~\ref{table1}, we arrive at $\sigma_{\mathrm{max}} \approx \sqrt{n_p}\times 10^{-5}$~nm. 
Given  $n_p  =  10^{12}$ for European XFEL \cite{Geloni2010}, $\sigma_{\mathrm{max}}\lesssim 10$~nm is required to achieve full nuclear inversion for all the presently considered isotopes. 

However, these requirements are easied by the large values that can be achieved for $R_c$ in the EWG.  Based on  Eq.~\eqref{eq9} in the Methods,   three requirements   determine the efficiency of a tapering structure: 
(i)  a few-nanometer x-ray focal spot, which is  defined by the radial FWHM of $R_c(r,z_{\mathrm{max}})$ at the focus and indicated by the leftward arrow in the left insets of Figs.~\ref{fig2}(b, c),
(ii) a long intrawaveguide Rayleigh length, which is  given by the axial FWHM of $R_c(0,z)$ at the focus and indicated by the horizontal arrows in the right insets of Figs.~\ref{fig2}(b, c),
and 
(iii) low photon loss.
The first two requirements maximize the nuclear excitation volume for a given input x-ray photon number, and the last criterion maximizes the  use of photons in focusing.  
As demonstrated in Fig.~\ref{fig2}, the  WG  enhances the free-space Rabi frequency with the maximum factor of $\sqrt{R_c \left( 0, z_f \right)}$ and therefore  reduces the required input XFEL photon number $n_p$ for achieving nuclear inversion with a factor of 28 for the EWG and of 14 for the CWG. The focusing spot radius and the Rayleigh length are respectively about 5~nm and 2~$\mu$m. 
This results in an active volume containing $10^6$ nuclei with solid-state particle number density. 
Moreover, the EWG can  achieve $T=0.8$ at the focus and so fulfil the condition (iii).

%%%%%%%%%%%%%%%%%%%%%%%%%%%%%%%%%%%%%%%%%%%%%
\begin{figure}[b]
\includegraphics[width=0.6\textwidth]{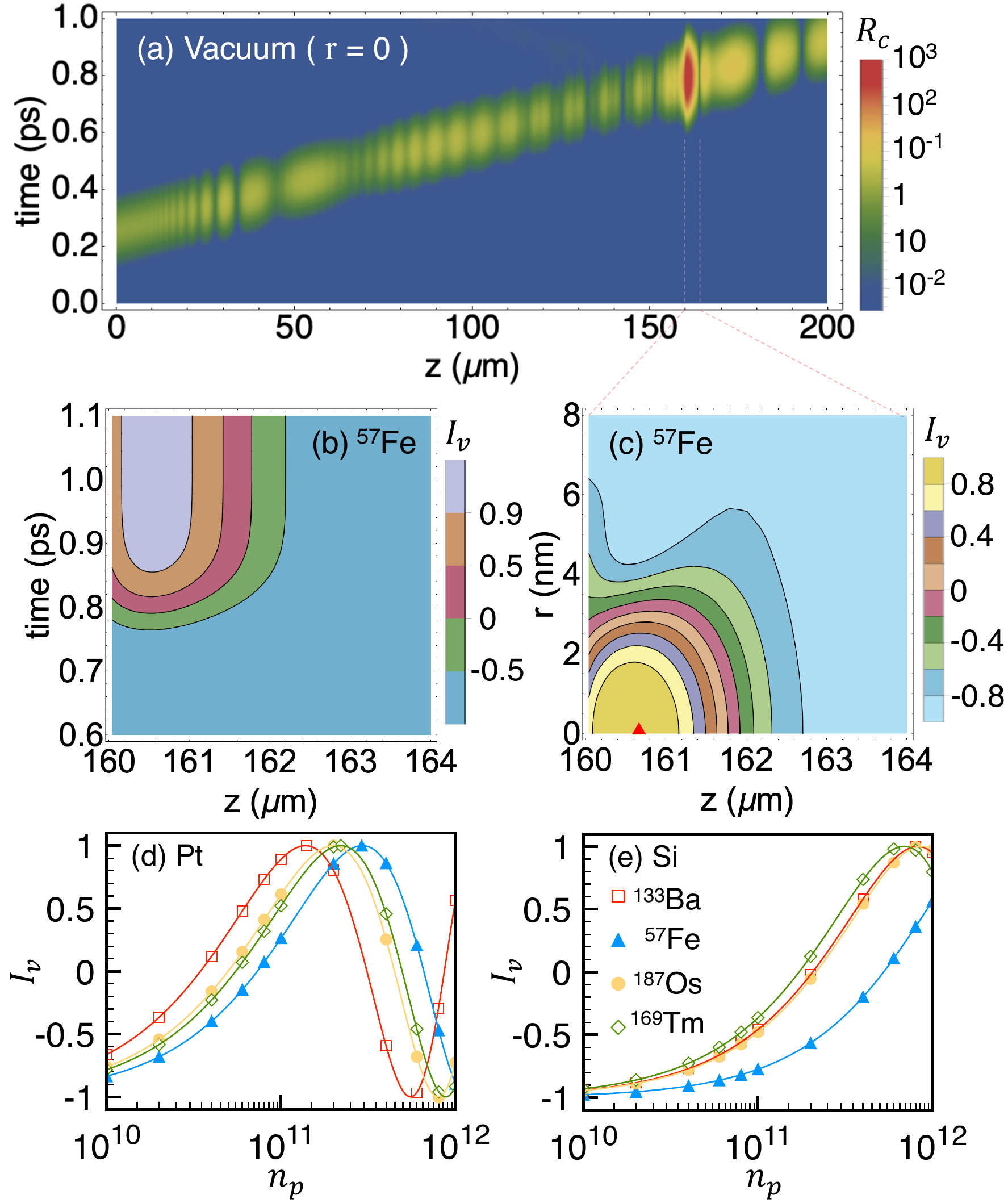}
\caption{\label{fig3} 
{\bf XFEL-pumped transient nuclear inversion in an elliptical waveguide.}
(a) Axial x-ray concentration $R_c\left( 0, z, t\right) $ as a function of time and axial coordinate $z$ for 14.4~keV x-ray propagation in a Pt-cladding EWG. 
(b) Time- and axial space-dynamics of nuclear population inversion $I_v$ around the EWG focus, and 
(c) spatial distribution of  $I_v\left( r, z\right) $ for $^{57}$Fe driven by a propagating XFEL with ($n_p$, $\sigma$ , $\tau$)  =  (2.88$\times 10^{11}$, 120~nm, 42.5~fs).
The nuclear population inversion maximum max$(I_v)$ for nuclear transitions in $^{133}$Ba (unfilled red square), $^{57}$Fe (filled blue triangle), $^{187}$Os (filled yellow circle), and $^{169}$Tm (unfilled greed diamond) is depicted as a function of the number of photons in the pulse   in a
(d) platinum-
(e) silicon-cladding EWG. We use the parameter set ($r_i$, $L_f$, $r_o$) = (150~nm, 160~$\mu$m, 15~nm) for all graphs.
}
\end{figure}
%%%%%%%%%%%%%%%%%%%%%%%%%%%%%%%%%%%%%%%%%%%%%

The full XFEL-nuclear dynamics is presented in  Fig.~\ref{fig3}. 
Figure~\ref{fig3}(a) demonstrates the propagation of an XFEL pulse through the EWG illustrated in Fig.~\ref{fig2}(c) where a $^{57}$Fe nuclear sample is embedded in $160\mu$m$\leq z\leq165 \mu$m. 
The axial $R_c\left( 0, z\right) $ shows that the XFEL intensity reaches the maximum when arriving at $z \approx 161\ \mu$m as also predicted by the steady-state solution in Fig.~\ref{fig2}(c).  
With XFEL parameters ($n_p$, $\sigma$ , $\tau$)  =  (2.88$\times 10^{11}$, 120~nm, 42.5~fs),  the corresponding $^{57}$Fe nuclear dynamics is investigated in Fig.~\ref{fig3}(b) which illustrates the axial nuclear inversion $I_v\left( 0, z, t\right) $. Figure~\ref{fig3}(b) shows that $^{57}$Fe is coherently excited by the coming x ray when the XFEL propagates through the WG neck. On the time scale of  100~fs XFEL FWHM, nuclei located near the focal point are fully inverted. As showed in Fig.~\ref{fig3}(c) which considers now the nuclear inversion at the end of the pulse, the XFEL leaves behind an  active volume of $\pi\times$3~nm$\times$3~nm$\times 1.7 \ \mu$m, containing $2.5\times 10^6$ inverted $^{57}$Fe nuclei out of  $4.1 \times 10^6$ totally embedded isotopes 
In Figs.~\ref{fig3}(d-e) we show the $n_p$-dependent maximum of the nuclear population inversion max$(I_v)$ for nuclear transitions in the isotopes $^{133}$Ba, $^{57}$Fe, $^{187}$Os and $^{169}$Tm  using Pt- and Si-cladding EWG, respectively. Each set of numerical results is fitted by Eq.~(\ref{eq9}) with a correspondingly colored solid line.

 Three conclusions can be drawn based on the results presented in Figs.~\ref{fig3}(d-e). 
First,
all four considered isotopes can be fully inverted around the focus in a Pt-cladding EWG in the input-photon-number range of $10^{11}< n_p < 10^{12}$. 
Significant inversion occurs also for photon numbers in the range  $10^{10}< n_p < 10^{11}$ in a Pt-cladding EWG for $^{133}$Ba, $^{187}$Os, and $^{169}$Tm. 
Population inversion in tapered waveguides appears to occur more efficiently and to require smaller photon numbers $n_p$ than in x-ray thin-film cavities according to theoretical predictions in Ref.~\cite{Heeg-arXiv2016}. 
The required x-ray photon number lies
within the reach of modern XFEL facilities \cite{LCLSBLs, SACLABLs, EuropeanXFELBLs, SwissFELBLs, PalFELBLs} and also of seeded XFEL facilities \cite{LCLS-II,Nam2021}. 
Second, the coherent x-ray pulse can drive a Rabi oscillation \cite{Buervenich2006,Palffy2008}. 
Finally, 
a comparison between 
Figs.~\ref{fig3}(d) and \ref{fig3}(e) demonstrates that the Pt-cladding EWG provides a better X-ray focusing than the silicon-cladding EWG. This behaviour  is due to the difference in index of refraction, which affects the critical reflection angle  and further leads to the enhanced x-ray focusing ability.

%%%%%%%%%%%%%%%%%%%%%%%%%%%%%%%%%%%%%%%%%%%%%%%%%%%%%%%%%%%%%%%%%%%%%%%%%%%%%%%%%%%%%%%%%%%%%%%%%%%%%%%%%%%%%%%%%%%%%%%%%%%%%%%
 
\begin{figure}[b]
\includegraphics[width=0.6\textwidth]{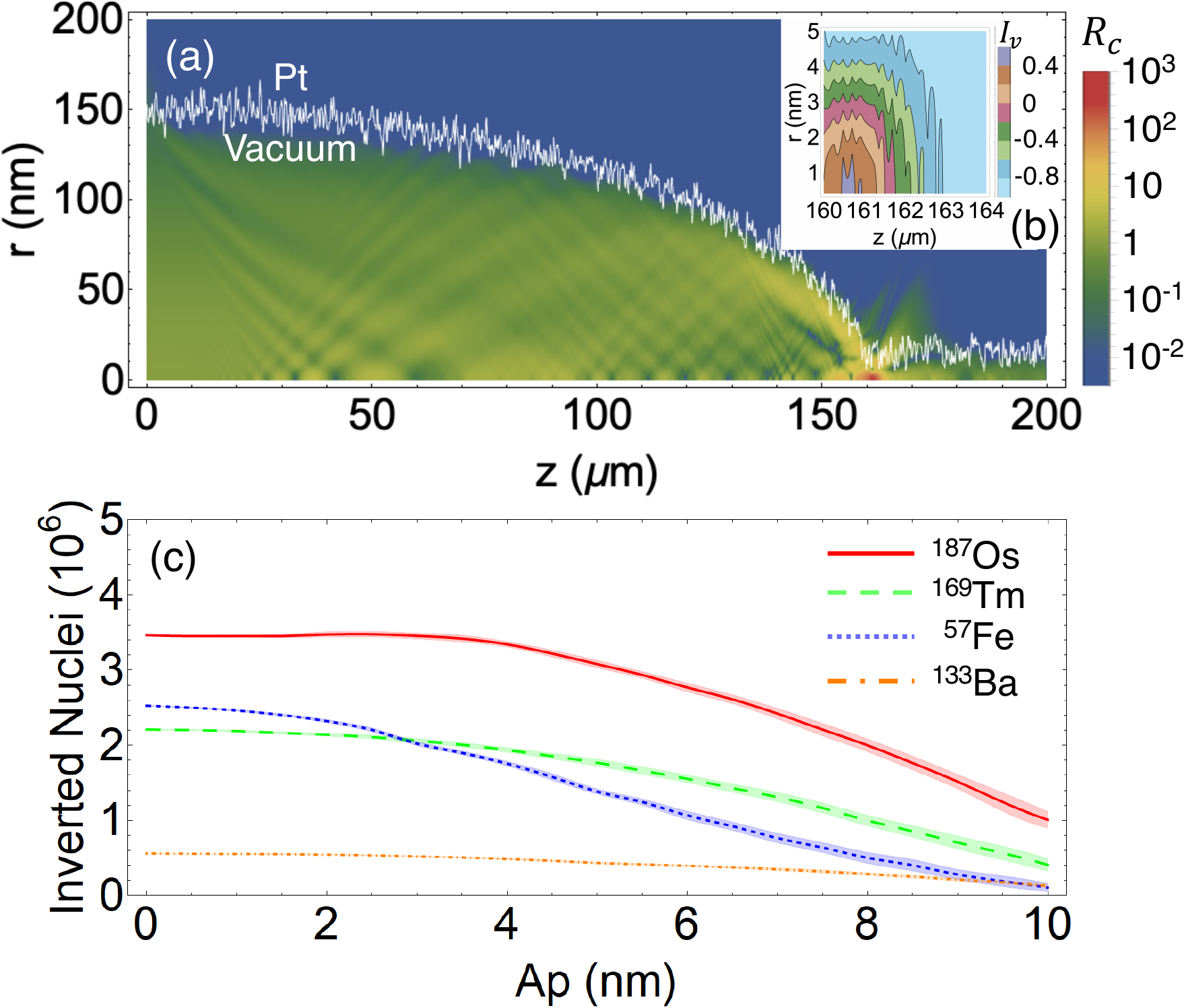}
\caption{\label{fig4} 
{\bf The effect of  waveguide roughness.}
(a) 14.4 keV EWG x-ray concentration distribution $R_c$ with a rough cladding interface of  ($r_i$, $L_f$, $r_o$,  $A_p$) = (150~nm, 160~$\mu$m, 15~nm, 5~nm).
The white line indicates the Pt-cladding-vacuum interface.
(b) the corresponding XFEL-pumped $^{57}$Fe $I_{v}\left( r, z \right) $ in the EWG  taking into account the surface interface roughness using the XFEL pulse parameters ($n_p$, $\sigma$ , $\tau$)  =  (2.88$\times 10^{11}$, 120~nm, 42.5~fs).
(c) $A_p$-dependent number of inverted $^{187}$Os (red solid line), $^{169}$Tm (green dashed line), $^{57}$Fe (blue dotted line), and $^{133}$Ba (orange dashed-dotted line) nuclei. We consider a WG with $L_f=160\ \mu$m. 
The shaded regions depict the error bars averaged over 100 realizations. 
}
\end{figure}

\noindent
{\bf The effect of  waveguide roughness.}
For a more realistic modelling of the WG, we investigate now the effect of surface imperfections at the vacuum-cladding interface. To this end we consider the cladding surface as ellipical with an additional white noise of amplitude $A_p$. Depending on the fabrication procedure, this amplitude can be on the order of several nanometers \cite{Osterhoff2009}. Figure~\ref{fig4}(a) illustrates an example of steady-state $R_c\left( r, z \right)$ using $A_p =$ 5~nm, where the maximum $R_c$ degrades from 800 [which was the case in Fig.~\ref{fig2}(c)] to 320. The direct consequence of the roughness is the reduction of nuclear excitation. 
Compared with Fig.~\ref{fig3}(c), our full calculation in Fig.~\ref{fig4}(b) shows
that the maximum $I_{v}$ becomes 0.4, and the inversion volume where $I_{v}>0$ shrinks by a factor of 0.6 resulting in $1.35\times 10^6$ inverted nuclei out of $2.5 \times 10^6$ totally embedded $^{57}$Fe isotopes. 
Note here that the active volume and therefore number of addressed nuclei changes when considering the WG roughness. 
In order to investigate the noise-amplitude-dependent total number of inverted nuclei, we numerically simulate  100 realizations, with random  sets of noise for  each amplitude value $A_p$ in Fig.~\ref{fig4}(c) considering the case of  $L_f = 160\ \mu$m. 
The results are illustrated for the four isotopes $^{187}$Os, $^{169}$Tm, $^{57}$Fe, and $^{133}$Ba. The shaded regions illustrate the error bars averaged over 100 realizations at each $A_p$. The quite small error bars reveal that the amount of inverted nuclei is very stable from noise to noise for a given $A_p$. The number of inverted nuclei sustains the increasing $A_p$ from 0 to 10~nm, and remains on the order of $10^6$. 

%
%%%%%%%%%%%%%%%%%%%%%%%%%%%%%%%%%%%%%%%%%%%%%%%%%%%%%%%%%%%

%%%%%%%%%%%%%%%%%%%%%%%%%%%%%%%%%%%%%%%%%%%%%%%%%%%%%%%%%%%
%

%%%%%%%%%%%%%%%%%%%%%%%%%%%%%%%%%%%%%%%%%%%%%%%%%%%%%%%%%%%%%%%%%%%%%%%%%%%%%%%%%%%%%%%%%%%%%%%%%%%%%%%%%%%%%%%%%%%%%%%%%%%%%%%

\begin{figure}[b]
\includegraphics[width=0.6\textwidth]{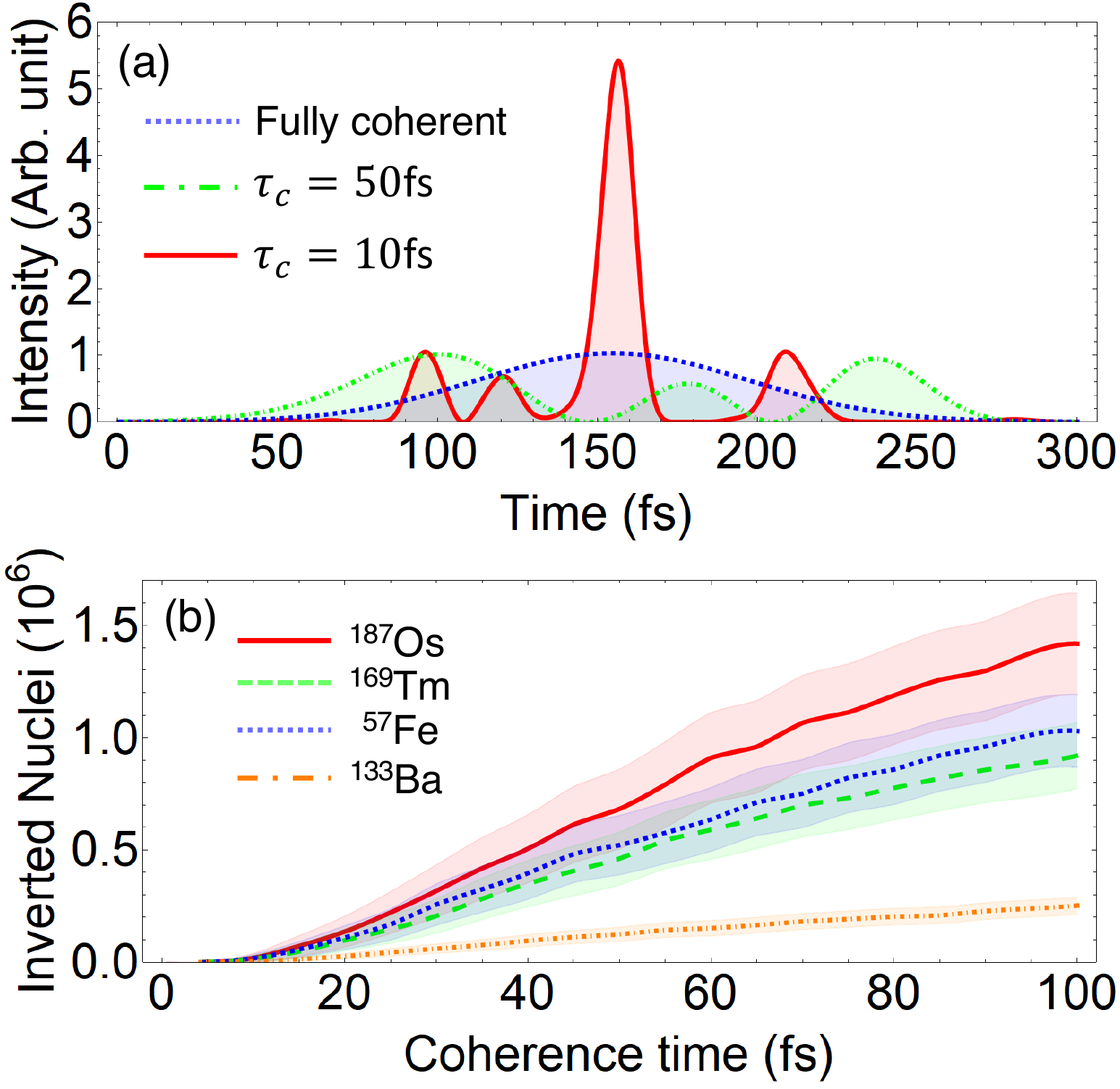}
\caption{\label{fig5} 
{\bf The effect of  XFEL coherence time.}
(a) XFEL intensity for  $\tau_c=50$~fs (green dashed dotted line), $\tau_c=10$~fs (red solid line), and considering full coherence (blue dashed line). 
The XFEL FWHM is 100fs for all cases.
(b) $\tau_c$-dependent number of inverted $^{187}$Os (red solid line), $^{169}$Tm (green dashed line), $^{57}$Fe (blue dotted line), and $^{133}$Ba  (orange dash-dotted line) nuclei for an EWG  with  ($r_i$, $L_f$, $r_o$, $A_p$) = (150~nm, 160~$\mu$m, 15~nm, 0). 
The shaded regions depict the error bars resulting from an average over 5000 SASE realizations, for illustration purposes scaled by a factor of 0.2.
}
\end{figure}

\noindent
{\bf The effect of XFEL coherence time.}
So far we have considered a completely coherent XFEL pulse. However, at present most XFEL facilities generate partially coherent x-rays  by self-amplified spontaneous emission (SASE) with good transverse coherence but rather short longitudinal coherence time $\tau_c$ of few tens of femtosecond \cite{Emma2010.NP}. Thus, in reality the time structure of a SASE XFEL pulse can be rather spiky. Using self-seeding \cite{Feldhaus1997},  the coherence time can be substantially improved, as already demonstrated at LCLS and SACLA  \cite{Amann2012,Inoue2019}.
We illustrate the effect of limited coherence on the XFEL pulse structure in Fig.~\ref{fig5}(a) which shows the pulse intensity for three different coherence times. The phase jumps randomly between neighbouring spikes  and degrades the ability of producing nuclear excitation. Moreover, the pulse energy fluctuates from shot to shot following the Gamma distribution \cite{Pfeifer2010}. In order to model these effects in calculating the nuclear population inversion, we use the partial-coherence method  to model SASE XFEL \cite{Pfeifer2010} combined with Eq.~\eqref{eq9} (see Methods). The results are illustrated in 
 Fig.~\ref{fig5}(b) for the four isotopes considered so far.  The shaded regions illustrate the error bars resulting from an averaging over 5000 SASE realizations at each $\tau_c$ with the EWG considered in Fig.\ref{fig2}(c). The error bars are 
scaled by a factor of 0.2 for better visualization. The quite large error bars reflect the SASE Gamma distribution. For all nuclear species, the amount of inverted nuclei reaches the  $10^4$ region  when $\tau_c > 10$~fs and approaches the result of fully coherent XFEL as increasing $\tau_c$. The statistics at $\tau_c = 50$~fs shows that  $10^6$   inverted $^{187}$Os, $^{169}$Tm, and $^{57}$Fe nuclei  are already produced by an average of  $\left\langle n_p \right\rangle \approx 2\times 10^{11}$ photons per pulse. Thus, strong nuclear inversion is potentially in reach also for just partially coherent XFEL pulses in an EWG.

%%%%%%%%%%%%%%%%%%%%%%%%%%%%%%%%%%%%%%%%%%%%%%%%%%%%%%%%%%%%%%%%%%%%%%%%%

%

%%%%%%%%%%%%%%%%%%%%%%%%%%%%%%%%%%%%%%%%%%%%%%%%%%%%%%%%%%%%%%%%%%%%%%%%%
\begin{widetext}
\begin{center}
\begin{table}[t]
\setlength{\tabcolsep}{6pt}
\vspace{-0.4cm}
\caption{\label{table1}
{\bf Nuclear and waveguide cladding parameters} \cite{nndc, lbl, ipmt}. For each nucleus with atomic mass number $A$  we present the nuclear transition energy $E_t$, the number density in solid-state $N$, the calculated Lamb-M\"ossbauer factor $f_{lm}$, 
the reduced nuclear transition probability $B(M1)$ in Weisskopf units (W.u.), the internal conversion coefficient $\alpha$ and the spontaneous radiative rate $\Gamma_s$ in MHz. The corresponding X-ray index of refraction is $n=1-\delta+i \beta$ for the respective nuclear target and the cladding materials are given in the last six columns.
}
\begin{tabular}{rrccccccccccc}
\hline
\hline
 
$^{A}X$ &   $E_{t}$ (keV) & $N $ ($10^{28}$/m$^{3}$) & $f_{lm}$ & $B(M1)$ (W.u.) & $\alpha$  & $\Gamma_s$ (MHz) & \multicolumn{3}{c}{ $\delta  (10^{-6})$} &  \multicolumn{3}{c}{$\beta (10^{-7})$ }
\\
\cline{8-10} \cline{11-13} 
& & & & & & & $^{A}X$&Pt & Si &$^{A}X$ &Pt & Si \\

\hline
$^{169}$Tm  & 8.410 & 3.32 & 0.84 & 0.0342 & 263 & 169 & 19.0 & 47.5  &6.94  &13.4 & 43.7 & 1.46\\

$^{187}$Os  & 9.756 & 7.15 & 0.96 & 0.0260 & 280 & 291 & 36.0 &34.7  & 5.14 & 24.6 & 25.7 & 0.81 \\

$^{133}$Ba  & 12.327& 1.54 & 0.38 & 0.0230 & 69.5& 99.0& 3.93 & 21.2 &3.21  & 3.10 & 28 &0.32 \\

$^{57}$Fe   & 14.413& 8.49 & 0.80 & 0.0078 & 8.65& 7.05& 7.44 & 16.1 & 2.34 &3.39  & 24.9 & 0.17\\

\hline
\hline
\end{tabular}
\end{table}
\end{center}
\end{widetext}
%%%%%%%%%%%%%%%%%%%%%%%%%%%%%%%%%%%%%%%%%%%%%%%%%%%%%%%%%%%%%%%%%%%%%%%%%
%

\section{Discussion}

Our results demonstrate that x-ray waveguides can increase the x-ray  intensity by up to three orders of magnitude and therefore substantially boost its efficiency to induce strong nuclear excitation. Combined with temporally coherent pulses,  prospects of nuclear population inversion finally come in reach at XFEL facilities. 
We emphasize that in this work we focus on the new aspect of waveguide-induced enhancement rather than  on collective nuclear dynamics previously investigated in the context of nuclear forward scattering \cite{Chumakov2018}.
Thus, we expect that our estimates of required photon number per XFEL pulse for achieving nuclear  inversion are rather conservative. A more sophisticated calculation including on equal footing both wavevuide field enhancement and nuclear collective effects deserves further investigation.

Experimentally, a stronger intensity carries with it more destructive power in form of heat load. This could be solved by  decreasing the off-resonant components of the incoming x rays. Possible implementations are
 for instance narrow-band monochromatization using a  $^{57}$FeBO$_3$ or $^{57}$Fe$_2$O$_3$ crystal \cite{Burck1980, Gerdau1985, Shvydko1996}.  This kind of nuclear monochromator has $<$10~neV bandwidth  and stretches the pulse duration to 100~ns \cite{Burck1980, Gerdau1985, Shvydko1996}. 
 Furthermore, the nuclear excitation detection requires  event-based electronics which record the scattered photons ideally in both time and energy. Such a so-called 2D-spectrum technique has been recently developed \cite{Goerttler2019, Heeg2021, Kuan2020}. It is anticipated that with higher count rates also the quantum properties of the scattered x-ray photons - for instance bunching or anti-bunching and photon correlations will become accessible to experiments.

The enhanced x-ray intensity in tapered waveguides will have a strong impact for the future development of x-ray quantum optics using the resonant interaction of x-rays with nuclear systems. Once demonstrated, the successful nuclear population control could be used for depleting nuclear isomers and thus releasing the energy stored therein. A classical example is the 2.4 MeV $^{93m}$Mo isomer with 6.8 hours halflife, which could be depleted by driving a 4.8 keV transition, well within the reach of XFEL facilities, upwards to a state which then directly decays within ns to the ground state \cite{Gunst2014,Wu2018}. Our calculations for this case show that, although nuclear population inversion requires higher photon per pulse  than presently available at XFEL facilities, an EWG enhances the achievable excitation  by a factor of approx.~200.

\section{Methods}
%%%%%%%%%%%%%%%%%%%%%%%%%%%%%%%%%%%%%%%%%%%%5

{\bf Optical-Bloch equations.}
The Liouville equation \eqref{Liouv} describing the coupling between x rays and nuclei can be written in terms of the density matrix elements   to yield the so-called optical Bloch equations \citep{Scully2006, Agarwal2012}
\begin{eqnarray}
\partial_{t}{\rho_{dd}} & = & \left( \frac{\alpha}{1+\alpha} \right) \Gamma_s\rho_{uu}+\frac{i}{2}\left(  \Omega^\ast \rho_{ud} - \Omega\rho_{ud}^\ast \right) ,\label{eq1}\\
\partial_{t}{\rho_{ud}} & = & -\left( \frac{\alpha}{1+\alpha} \right) \frac{\Gamma_s}{2} \rho_{ud}-\frac{i}{2}\Omega\left(   \rho_{uu} - \rho_{dd} \right) ,\label{eq2}\\
\partial_{t}{\rho_{uu}} & = & -\left( \frac{\alpha}{1+\alpha} \right) \Gamma_s\rho_{uu}-\frac{i}{2}\left(  \Omega^\ast \rho_{ud} - \Omega\rho_{ud}^\ast \right) ,\label{eq3}
\end{eqnarray}
where the subscripts $\left(  u, d\right) = \left(  3, 1\right) $ or $\left(  4, 2\right) $ label the nuclear states as illustrated in Fig.~\ref{fig1}(c). In Eq.~(\ref{eq1}-\ref{eq3}), the incoherent decay of the nuclear excitation is attributed to 
the internal conversion rate $\frac{\alpha}{1+\alpha}\Gamma_s$, and the  WG mode-nuclei coupling is described by the coherent process. We solve the optical Bloch equations together with the wave equation for the propagation of the x-ray electric field \eqref{eq4} using the initial conditions  $\rho_{33}\left( r, z, 0\right)=\rho_{44}\left( r, z, 0\right)=\rho_{31}\left( r, z, 0\right)=\rho_{42}\left( r, z, 0\right) = E\left( r, z, 0\right) = 0$, $\rho_{11}\left( r, z, 0\right)=\rho_{22}\left( r, z, 0\right)= 1/2$. The boundary condition \cite{Palffy2008, Liao2011, Liao2013} at $z = 0$ is 
\begin{equation}\label{eq5}
E\left( r, 0, t\right)  = \sqrt{\frac{n_p E_t \sqrt{2}}{c\epsilon_0 \sigma^2\tau\sqrt{\pi^3}}}
\exp\left(  -\frac{r^2}{\sigma^2} - \frac{\left( t -t_c\right) ^2}{4\tau^2} \right) . 
\end{equation}
Here , $\epsilon_0$  is the vacuum permittivity, $t_c=250$fs is the XFEL peak time, and $E_t$ the nuclear transition energy, respectively. Furthermore, 
$n_p$, $\sigma$, and $\tau$ are the photon number per pulse,  beam waist, and pulse duration of the input XFEL pulse, respectively. 
We note that, together with the  dispersion relation in the wave Eq.~(\ref{eq4}),  the Gaussian pulse modeling in Eq.~(\ref{eq5}) is more realistic than the originally used effective intensity approach \cite{Buervenich2006} 
and renders larger excited state populations $\rho_{uu}$. Numerically, in Eq.~\eqref{eq4} the  contribution of the nuclear coherences term on the right-hand side is dominated by the waveguide terms.

Following Ref.~\cite{Scully2006,Palffy2008,Liao2011,Liao2012a}, we define the Rabi frequency 
$\Omega =\Pi E/\hbar$, where
\begin{equation}\label{eq6}
\Pi =
\frac{4a\mu}{3}
\sqrt{1.79\left( 2I_u+1\right)\pi  B\left(M1\right)  } \ ,
\end{equation}
$\mu = 0.105 e\cdot$fm is the nuclear magneton,  $a=\sqrt{2/3}$ is the corresponding Clebsch-Gordan coefficient, and 
$B\left(M1\right)$ is the reduced transition probability for the magnetic dipole nuclear transition. 
All nuclear and WG cladding parameters used in our calculations are listed in Table~\ref{table1}.
For simplicity, we define 
%the total excited state population $\rho_{ee}=\rho_{33}+\rho_{44}$ and 
the nuclear inversion $I_v = \rho_{33}+\rho_{44} -\rho_{11}-\rho_{22}$ \cite{Buervenich2006}.
%We numerically solve the above OBE by using the xxx method.
%
When obtaining the x-ray concentration distribution $R_c$ via a steady-state numerical calculation, i.e., solving Eq.~(\ref{eq4}) without nuclear coherence and temporal derivative terms \cite{Pfeiffer2002, Fuhse2006}, one  can calculate the XFEL pulse area in a typical two-level model and obtain the nuclear inversion
\begin{equation}\label{eq9} 
I_v\left( r, z\right)  = -\cos\left[ 2\frac{\Pi}{\sigma\hbar}\sqrt{\frac{\tau n_p T R_c \left( r, z\right) E_t \sqrt{2}}{c \epsilon_0 \sqrt{\pi}}}\right] .
\end{equation}

\noindent
{\bf Coherence time for SASE XFEL.}
%%%%%%%%%%%%%%%%%%%%%%%%%%%%%%%%%%%%%%%%%%%%%%%%%%%%%
We use the partial-coherence method  to numerically simulate the  SASE XFEL pulse structure \cite{Pfeifer2010}. The pulse energy 
$P = 2\pi c\epsilon_0 \int_{-\infty}^\infty \int_0^\infty \vert E\left( r, 0, t\right)\vert^2 rdrdt$ 
follows a Gamma distribution:
\begin{equation}
\Gamma\left( P\right) =\frac{M^M}{\Gamma\left( M \right) } \left( \frac{P}{\left\langle P\right\rangle }\right)^{M-1}\frac{1}{\left\langle P\right\rangle }\exp\left( -M\frac{P}{\left\langle P\right\rangle }\right),  
\end{equation}
with the mode number 
\begin{equation}
M=\frac{\left\langle P\right\rangle ^2}{\left\langle \left( P-\left\langle P\right\rangle \right)^2 \right\rangle }.
\end{equation}
In order to investigate the effect of SASE XFEL as demonstrated in Fig.~\ref{fig5},  we obtain the  lowest average pulse energy $\left\langle P\right\rangle$ for achieving the full nuclear inversion $I_v=1$ in Eq.~(\ref{eq9}) with the maximum EWG x-ray concentration $R_c$:
\begin{equation}
\left\langle P\right\rangle = \left\langle n_p \right\rangle  E_t = \frac{ c \epsilon_0 \pi^{3/2} \sigma^2\hbar^2 }{4\sqrt{2} \tau  \Pi^2 \mathrm{max}\left( R_c \right) }.
\end{equation} 
For each isotope listed in Table~\ref{table1}, $\left\langle P\right\rangle / E_t = \left\langle n_p \right\rangle \approx 2\times 10^{11}$ photons per pulse. The number of inverted nuclei can be calculated as
\begin{equation}
\int_{I_v>0}N\left( \vec{r} \right) \left[  \rho_{33}\left( \vec{r} \right)+\rho_{44}\left( \vec{r} \right) \right] d^3 \vec{r}
\end{equation}
for Figs.~\ref{fig4}(c) and \ref{fig5}(b).

\section{Acknowledgements}
Y.-H.~C., P.-H.~L, G.-Y.~W., and W.-T.~L. are supported by the Ministry of Science and Technology, Taiwan (Grant No. MOST 107-2112-M-008-007-MY3 \& MOST 109-2639-M-007-002-ASP). 
%W.-T.~L. is also supported by the National Center for Theoretical Sciences, Taiwan. 
AP gratefully acknowledges support from the Heisenberg Program of the Deutsche  Forschungsgemeinschaft (DFG). 

\section{Author contributions}
Y.-H.~C. derived the physics model and developed the computational code.
Y.-H.~C., P.-H.~L, and G.-Y.~W. performed the numerical calculation.
A. P. derived the nuclear condensed matter input of the calculation. W.-T.~L. conceived the idea and conducted the project. All the authors discussed the results and wrote the manuscript.

%%%%%%%%%%%%%%%%%%%%%%%%%%%%%%%%%%%%%%%%%%%%%%%%%%%%%%%%%%%%%%%%%%%%%%%%%

%%%%%%%%%%%%%%%%%%%%%%%%%%%%%%%%%%%%%%%%%%%%%%%%%%%%%%%%%%%%%%%%%%%%%%%%%

%%%%%%%%%%%%%%%%%%%%%%%%%%%%%%%%%%%%%%%%%%%%%%%%%%%%%%%%%%%%%%%%%%%%%%%%%

%%%%%%%%%%%%%%%%%%%%%%%%%%%%%%%%%%%%%%%%%%%%%%%%%%%%%%%%%%%%%%%%%%%%%%%%%%%%%%%%
\end{document}